\documentclass[english]{amsart}
\usepackage[T1]{fontenc}
\usepackage[latin9]{inputenc}
\usepackage{amsmath}
\usepackage{graphicx}
\usepackage{longtable,booktabs}
\usepackage{graphicx,grffile}

\def\re#1{(\ref{#1})}   
\providecommand{\tabularnewline}{\\}

\makeatother

\usepackage{babel}
\begin{document}

\title{Seismic noise measures for underground gravitational wave detectors}

\author{Somlai L.$^{1,2}$ and Gráczer Z.$^{3}$, Lévai P.$^{1}$, Vasúth M.$^{1}$, \\Wéber Z.$^{3}$, Ván P.$^{1,4,5}$}

\address{$^{1}$MTA Wigner Research Centre for Physics, Institute of Particle and Nuclear Physics, 1121 Budapest, Konkoly Thege Miklós út 29-33, Hungary;\\
$^{2}$University of Pécs, Faculty of Sciences, H-7624 Pécs, Ifjúság út 6, Hungary;\\
$^{3}$MTA Research Centre for Astronomy and Earth Sciences, Geodetic and Geophysical Institute, H-9400, Sopron, Csatkai E. u. 6-8, Hungary;\\
$^{4}$Budapest University of Technology and Economics, Department of Energy Engineering, 1111 Budapest, Bertalan Lajos u. 4-6, Hungary;\\
$^{5}$Montavid Thermodynamic Research Group, Budapest, Hungary.}

\begin{abstract}
The selection of sites for underground gravitational wave detectors based on spectral and cumulative characterisation of the low frequency seismic noise. The evaluation of the collected long term seismological data in the Mátra Gravitational and Geophysical Laboratory revealed several drawbacks of the previously established characteristics. Here we demonstrate the problematic aspects of the recent measures and suggest more robust and more reliable methodology. In particular, we show, that the mode of the data is noisy, sensitive to the discretization and intrinsic averaging, and the $rms_{2Hz}$ is burdened by irrelevant information and not adapted to the technological changes. Therefore the use of median of the data instead of the mode and also the modification of the frequency limits of the $rms$ is preferable.
\end{abstract}

\maketitle

\section{Introduction}

The improved sensitivity of future third generation gravitational wave (GW) detectors requires various technological developments. One of the plans is to optimize the facility for underground operation, in order to reduce the noise between the frequencies from 1~Hz to 10~Hz. According to the related sensitivity calculations the seismic and Newtonian noises represent the most important noise contributions in this frequency range \cite{BraEta10p,Har15a}. During the preparatory studies of the so-called Einstein Telescope (ET), the European initiative, several short term seismic measurements were performed in various locations \cite{ETdes11r,BekEta15a,Bek13t}. Based on these studies two performance measures were established: a spectral and a cumulative one. According to the spectral recommendation the average horizontal acceleration Amplitude Spectral Density should be smaller than the $A_{BF}$ limit,

\begin{equation}
A_{BF}=2\cdot10^{-8}\frac{m/s^{2}}{\sqrt{Hz}},\,\,\:\text{in the region}\,\, 1Hz\leq f\leq 10Hz.\label{eq:ETReqBF}
\end{equation}

This is the so-called Black Forest line
, named after one of the investigated sites. This spectral criterion corresponds to a cumulative value, the square root of the displacement Power Spectral Density integrated from the Nyquist frequency down to $2~Hz$, this is the $rms_{2Hz}$ and its value for the Black Forest line is $0.1\,nm$.

In the ET survey the three best sites that fulfilled these requirements are the LSC Canfranc laboratory in Spain ($rms_{2Hz}=0.070\,nm$), the Sos Enattos mine in Sardinia, Italy ($0.077\,nm$) and the Gy\"ongy\"osoroszi mine in Hungary ($0.082\,nm$ and  $0.12\,nm$ in depths $400\,m$ and $70\,m$ respectively). The data collection was performed up to a week at most in the various sites. In spite of the similar cumulative $rms_{2Hz}$ values, the spectra of these sites is far from being uniform: the contributions of civilization noise, oceanic and sea waves appear in different frequency ranges and with different weights. 

The M\'atra Gravitational and Geophysical Laboratory (MGGL) has been operating since March 2016 with the purpose to evaluate and survey the M\'atra mountain range as a possible ET candidate site. The primary goal of the laboratory is to collect seismic noise data for long period and evaluate them from the points of view of ET \cite{BarEta17a}. The laboratory is located at the coordinates (399 MAMSL, 47$^{\circ}$52' 42.10178\textquotedbl{}, 19$^{\circ}$51' 57.77392\textquotedbl{} OGPSH 2007 (ETRS89)), along a horizontal tunnel of the mine, 1280~m from the entrance, 88~m depth from the surface. It is situated near to the less deep location of the above mentioned former short term measurements and it is prepared for long term data collection in a telemetric operation mode. In the laboratory a Guralp CMG-3T seismometer (hereafter referred as ET1H) was installed and has been operating continuously except shut downs which happens at strongly interfering mine activities (e.g. explosions). There is an ongoing reclamation activity in the mine and therefore the human activity is not negligible in recent years. The regular operation of the mine railway, the continuously working large water pumps in the vicinity of the laboratory and the related technical service and construction activities are producing industrial noise. These instruments will not be present in the future, especially during gravitational wave detection.


The ET related analysis of long term noise data revealed some particular aspects, that are not apparent in short term measurements, and could influence the operational conditions and detection possibilities of GWs in an underground location. Therefore we need to expose these effects for the optimal operation of the detector facilities. These are in particular the presence of various short term seismic disturbances with large amplitudes and the methodology of long term data evaluation.  

The short time, large amplitude disturbances are unpredictable, unavoidable and must be left out to obtain reliable estimation of the average low noise level. However, any particular truncation or cutting process generates biases on the spectral and also the cumulative noise measures. {\em To avoid these biases we suggest to use the percentiles of the complete data.} The percentiles select the highest and lowest values, this selection is relative, and based on the intrinsic feature of the data set. 

Any long term analysis and the evaluation of spectra and $rms$ may require some intermediate averaging over the basic averaging length of the Fourier transformation. For this purpose here we suggest two different averaging steps: 
\begin{enumerate}
	\item[(a)] calculate short time averages (STA) to get manageable size of the data sets and to use the optimal time-scales of the planned detector. 
	\item[(b)] calculate intermediate -- for whole day, night or working periods i.e. natural periodicity of the data -- percentiles and analyse the averages of them to study daily, annual, etc. variations. In the following we will call this intermediate or long time averaging as intrinsic averages (INA).
\end{enumerate}
In particular  the averaged daily percentiles of the complete data set can be used to estimate the spectral and cumulative variation of the data and the averaged daily median -- the $50th$ percentile -- values for the comparison with the Black Forest line. If the data collection period is longer than half year, then it is practical to use INA.

The paper is organized as follows. First we shortly survey the evaluation procedure including its pitfalls, like the usage of mode for $rms$ calculation. Then we analyse the effect of averaging on spectral and cumulative measures established by Beker {\it et al} \cite{BekEta15a}. After that the utility of INA is studied and we examine $rms$ values with different frequency interval. Finally we conclude our experiences of calculation process and measurements and suggest further quantities to compare sites.

\section{Data and data analysis \label{sec:_seismo_measurements}}

The new seismological data for our recent analysis were collected by a Guralp CMG 3T low noise, broadband seismometer, which is sensitive to ground vibrations with flat velocity response in the frequency range 0,008-50Hz. The self noise of the seismometer is below the New Low Noise Model  of Peterson in the region $0.02\,Hz$ to $10\,Hz$ \cite{Guralpmanual}. 
In this paper we study data collected by one instrument (ET1H). This station was permanently installed in the MGGL. The seismometer is deployed on a concrete pier which is connected to the bedrock. Between the pier and the seismometer a granite plate has been placed. The data collection period for ET1H has been started on 2016-03-01. In this paper we focus on methodology restricting the studied data period from 2017-01-01 until 2017-12-15 (349 days).


In our analysis we followed the data processing method of \cite{BekEta15a}, e.g. the so-called Nuttal-window was applied with $3/4$ overlap. In this section we recall the basic definitions. The Power Spectral Density (PSD) for the velocity is defined as
\begin{equation}
P^{(v)}=\frac{2}{f_{s}\cdot N\cdot W}\left|V_{k}\right|^{2},
\end{equation}
where $f_{s}$ is the sampling rate, $N$ is the length of the analysed data sample, and $W=\frac{1}{N}\sum_{n=1}^{N}w[n]^{2}$ with the Nuttall window function $w[n]$. The coefficients $V_{k}=F(w[n]\cdot(v[n]-\langle v\rangle)$, represent the Fourier transform $F$ of the deviation of raw velocity data $v[n]$ from its average value $\langle v\rangle$. In our analysis PSDs were calculated with $50\,s$ data samples. The choice of this sample length for Fourier transformation is a compromise between the frequency resolution of the spectra and the detectability of short noisy events. The resulted $0.02Hz$ resolution seems to be reasonably fine and we can reliably identify less than a second long seismic events.  We did not use the advantage of fast Fourier algorithm on the expense of increasing the lowest frequency value\footnote{Other instruments, the Trillium seismometer of the previous study, work with $128~Hz$ sampling rate. Then the $128\,s$ interval is convenient for fast Fourier calculation, but hourly or daily spectra require truncations.}. Before further processing, raw data were highpass-filtered with $f_{HP}=0.02\,Hz$. 

Our STA is chosen to be $300\,s$. As we have mentioned above, the basic Fourier length is influenced by the sampling rate of the instrument. On the other hand for long term data the analysis can be easily adapted to the natural human and industrial noise periods. In the previous studies STA was $1800\,s$, which is natural with the basic $128\,s$ Fourier length, considering the overlap. With our choice of STA, the comparison of the two analysis is with minimal bias, simply because $6\times 300\,s\,=\,1800\,s\approx14\times128\,s$.

The Amplitude Spectral Density (ASD) for the velocity can be calculated from PSD via $A^{(v)}=\sqrt{P^{(v)}}$. Both amplitude and power spectral densities can be expressed also as either acceleration ($a$) or displacement ($d$) by multiplying or dividing by $\omega=2\cdot\pi\cdot f$ or the square of it respectively. For example, $A^{(d)}=A^{(v)}/\omega$. Therefore the mentioned ET comparison level, \re{eq:ETReqBF}, can be transferred easily to other spectral densities, e.g. for the Black Forest line:
\begin{equation}
P^{(a)}_{BF} = (A^{(a)}_{BF})^2 = 4\cdot10^{-16}\frac{m^{2}/s^{4}}{Hz}\,\,\,\text{or}\,\,\,
P^{(d)}_{BF}=\left(\frac{A_{BF}}{\omega^2}\right)^2 = 4\cdot10^{-16} \omega^{-4} \frac{m^{2}}{Hz}
\label{eq:ETReqv}\end{equation}

\begin{figure}
\centering
\includegraphics[scale=0.5]{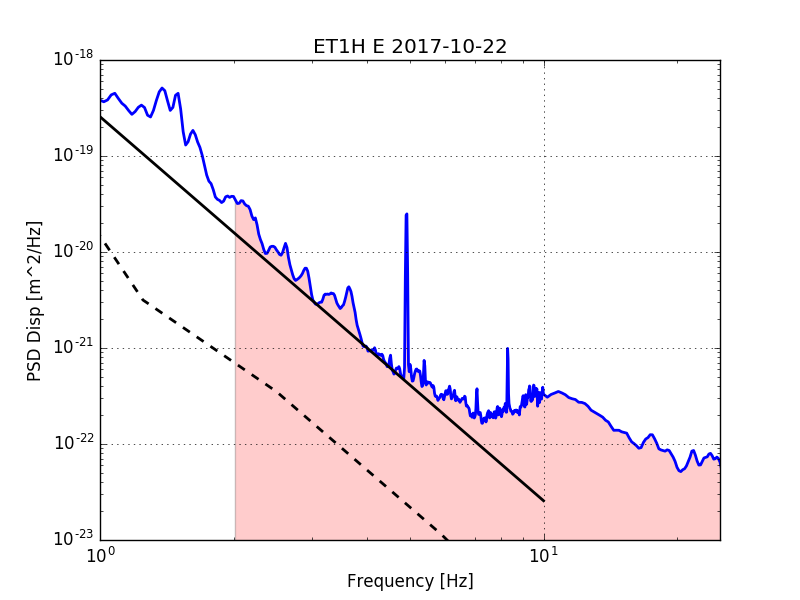}
\caption{\label{fig:rms-illusztration} Illustration of $rms$ at a displacement PSD spectrum (blue line) together with the Black Forest line (solid black) and the New Low Noise Model of Peterson (dashed black). The filled area represents $rms_{2Hz}^2$.}
\end{figure}

It is convenient to characterize sites in terms of acceleration ASD spectra and its variation and also by displacement $rms$ as a single cumulative property. The displacement $rms$ is the square root of the integral of displacement PSD between two frequency values
\begin{equation}
rms^{(d)}=\sqrt{\frac{1}{T}\sum_{k=l}^{N/2+1}P_{k}^{(d)}},
\end{equation}
where $l$ is the cutoff index, $T=\frac{N}{f_{s}}$. The usual choice is $2\,Hz$ for comparing ET candidate sites \cite{BekEta15a}. The displacement power spectral density of the daily average of 2017-10-22 of ET1H station, East direction is shown on Figure \ref{fig:rms-illusztration}. The Black Forest line is the solid straight line, the New Low Noise Model of Peterson (NLNM) \cite{Pet93a} is the dashed one. The $rms_{2Hz}$ is the square of the area of the shaded region. 

For the comparison of various spectra of ET sites it is worth to show the Black Forest line, Eq. (\ref{eq:ETReqBF}) and to recall the corresponding $rms^{(d)}=0.1nm$ value at 2Hz:
\begin{gather}
rms_{2Hz}^{(d)}=\sqrt{\int_{2}^{f_{s}/2}P^{(a)}_{BF}\frac{1}{\left(2\pi\right)^{4}}\frac{1}{f^{4}}df}=
\frac{A_{BF}}{\left(2\pi\right)^{2}}\sqrt{\int_{2}^{f_{s}/2}\frac{1}{f^{4}}df}\approx\nonumber \\\frac{A_{BF}}{\left(2\pi\right)^{2}}\sqrt{\left[f^{-3}/(-3)\right]_{2}^{\infty}}\approx0.1nm,\label{eq:rms2hz}
\end{gather}
where we considered that the displacement PSD values decrease significantly at higher frequencies and expanded the domain of BF line to the infinity with the same value. See Figure \ref{fig:rms-illusztration} as an illustration, where it is obvious that in some cases higher frequencies can contribute significantly to the $rms$ value. For the particular data shown in Figure \ref{fig:rms-illusztration} the $rms_{2Hz}=0.209\,nm$, the $rms_{2-10Hz}=0.144\,nm$ and their ratio is $69.7\%$. 

In the following only displacement $rms$ will be used so the $(d)$ superscript is omitted. Furthermore, two more $rms$ will be considered:
the $rms_{2-10Hz}$ and the $rms_{1-10Hz}$. For the Black Forest line they are:
\begin{eqnarray}
rms_{2-10Hz}&=&\sqrt{\int_{2}^{10}P^{(a)}_{BF}\frac{1}{\left(2\pi\right)^{4}}\frac{1}{f^{4}}df}\approx0.1nm, \\
rms_{1-10Hz}&=&\sqrt{\int_{1}^{10}P^{(a)}_{BF}\frac{1}{\left(2\pi\right)^{4}}\frac{1}{f^{4}}df}\approx0.29nm.\label{eq:rms1-10hz}
\end{eqnarray}
In Section \ref{frequency-range-and-rms} we will show how both values can specify new information about the site.

\section{Mode vs. median\label{subsec:Mode-vs.-median}}

\begin{figure}
	\centering
	\includegraphics[width=0.95\textwidth]{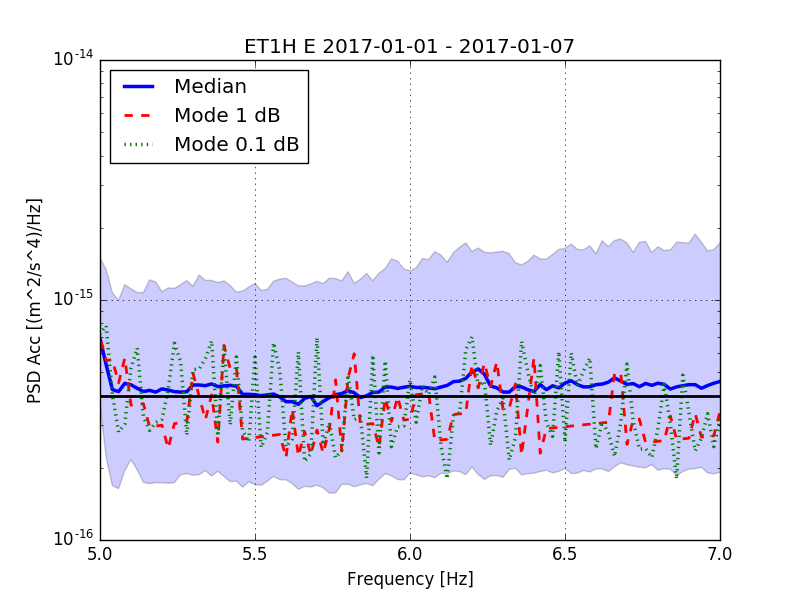}
	\caption{\label{fig:MedianVsModusPSDACC} In this figure the median (solid blue line)  and the modes (dashed and dotted lines)  with $1\,dB$ and $0.1\,dB$ bins are compared. The solid black horizontal line represents the Black Forest line and the blue area indicates the 10th-90th percentiles.}
\end{figure}

In Beker {\em et al}. \cite{BekEta15a} the mode of the STA was used to characterize the typical noise level. Here we show that for long term data analysis the mode strongly depends on the discretization of the spectrum. In order to illustrate the differences, we use the same method as Beker \cite{BekEta15a} to determine modes. Only 7 days of data, between 2017-01-01 and 2017-01-07, was chosen for the recent analysis. The modes are shown together with the $10th$, $50th$  and $90th$ percentiles of the half-hour averages on Figure \ref{fig:MedianVsModusPSDACC} between $5\,Hz$ and $7\,Hz$, with $1\,dB$ and $0.1\,dB$ bins. It is clear the fluctuation of the mode is discretisation dependent and larger than the fluctuation of the median. It is remarkable that the $rms_{5-7Hz}$-s are $0.0218nm$, $0.0198nm$ and $0.0211nm$ for the median, mode $1\,dB$ and $0.1\,dB$ respectively.

\begin{figure}
	\centering
	\includegraphics[width=0.95\textwidth]{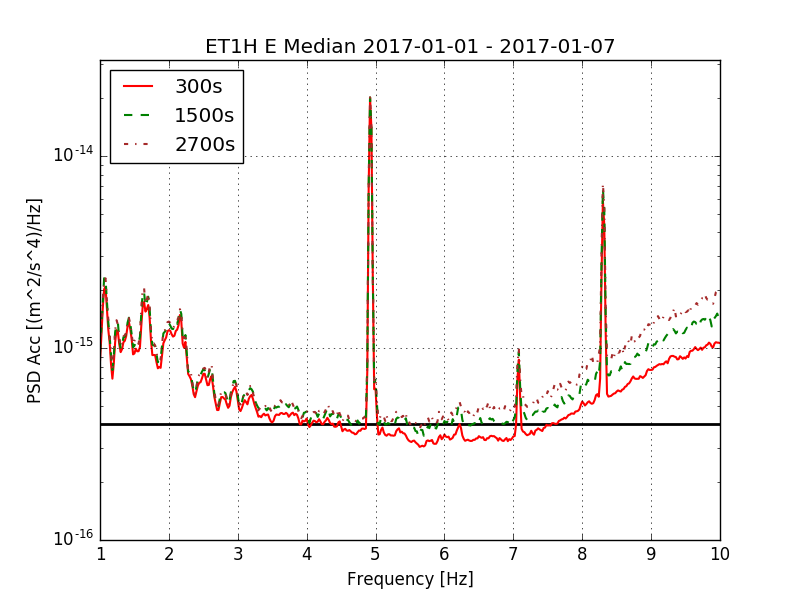}
	\includegraphics[width=0.95\textwidth]{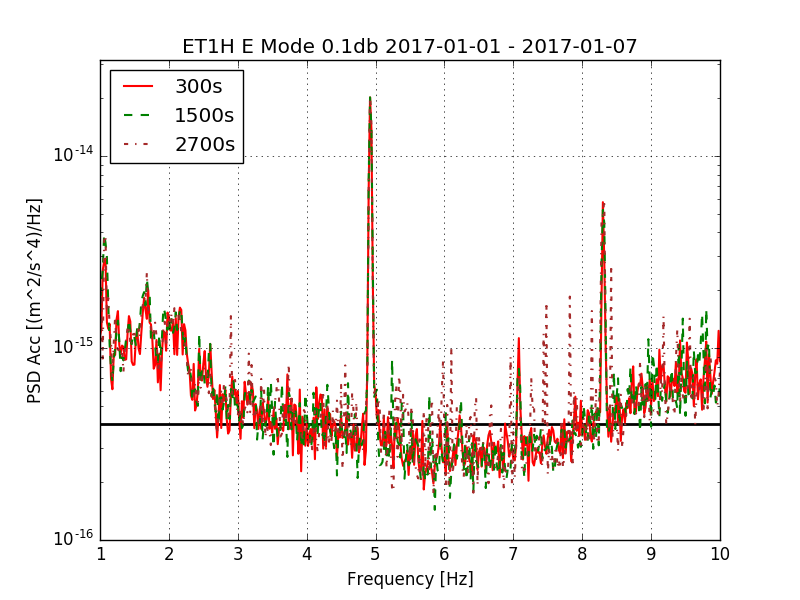}
	\caption{\label{fig:MedianVsModeSTAdepend} Upper figure displays the median with different short time averaging lengths. Lower figure shows the mode of the same data  with the same different STA lengths. The Black Forest line is solid black. The data is from the first week of January in 2017. }
\end{figure}

Our next step is to illustrate the advantage of median when considering different short time averaging (STA) lengths. A "well-behaving" characterization is expected to keep its profile for different STAs, in order to avoid process dependent artifacts. As it was mentioned, the STA is 300s in our case. In Beker's site selection study \cite{BekEta15a} approximately half-hour ($1800s$) STA was chosen. The differences between the expected values are illustrated in Figure  \ref{fig:MedianVsModeSTAdepend}.  The mode is noisier than the median. The median is slightly increasing above $5~Hz$ with increasing STA.

For the median there is no need for noise level discretization and it is not sensitive to noise level distribution. In general it is a more stable quantity. The following simple example demonstrates this. Let us consider the half-hour PSD-s calculated for one-week interval. Then we have about 350 samples. Then a PSD bin with 0.1dB, and an 10dB difference between the 10th, 90th percentiles, we obtain 100 bins for calculating the mode. With a sufficiently uniform noise level distribution at a given frequency only 4 PSD values could define the mode. Then it is understandable, hat with several noise peaks with varying strength the mode can fluctuate violently. On the other hand the median characterizes the best/worst $50\,\%$ of the data, it is not sensitive of the form of the distribution and does not require power discretisation. Therefore, in the following analysis {\it the use of median is preferred}. 

\section{The effect of intrinsic averaging}

To study long term -- annual and seasonal -- seismic noise variation and investigate site properties for the planned detector the use of intrinsic averages can also be necessary. Considering one year of data implies $365\,days\times288\,STA \approx 100\,000$ data point so the $90th$ percentile is defined by the $10\,000$ worst STAs. It could be a problem that one has not get any information about the density distribution: the $90th$ percentile is defined by just few noisy months or by three hours every day. Therefore intrinsic averaging (INA) is suggested to handle this difficulty and it can also be use to optimize the process whether discretization is omitted or not. The natural periodicity of the noise data indicates the use of daily averaging.

\begin{figure}
	\centering
	\includegraphics[width=0.95\textwidth]{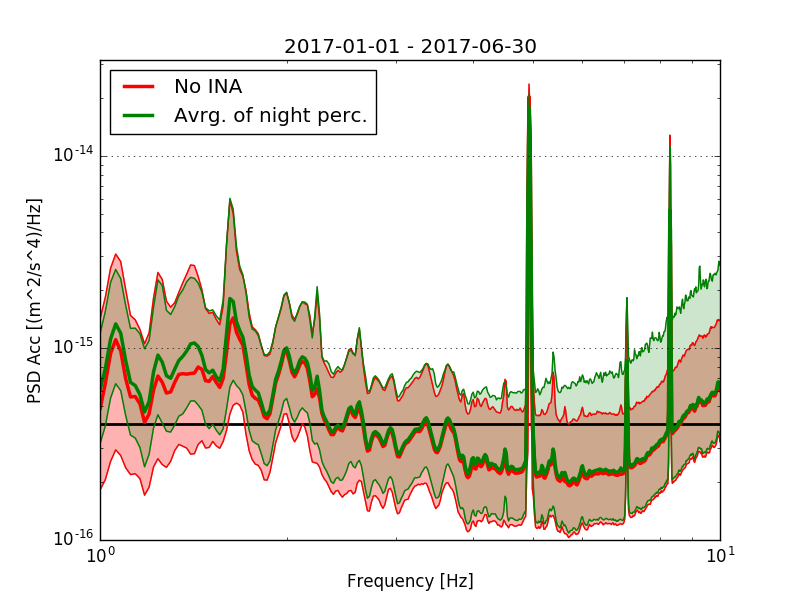}
	\caption{\label{fig:INA-diff-night1}Here the differences of spectra with and without intrinsic averaging (INA) are shown. Red curves belong to no INA and green ones to the averages of night percentiles -- averages of 10th, and 90th percentiles are the upper and lower limits of the shaded area. The median is shown with green and red lines in the middle.}  	
\end{figure}

To illustrate it, we defined night period (00:00 - 2:00 and 20:00 - 24:00 UTC) -- in order to reduce the effect of human activity and focus on the noise changes --  and calculated the percentiles with and without INA in Figure \ref{fig:INA-diff-night1}. As it can be seen the medians have almost the same values for the whole interval, but either $10th$ or $90th$ percentiles show slightly different properties of the site. In general the use of intrinsic averaging shows small differences when compared to the evaluation without INA. The spectrum is slightly noisier with INA, therefore we cannot underestimate the noise level using that. On the other hand for large amount of data the analysis and the calculations are more convenient with INA.

\section{The frequency range and cumulative statistics\label{frequency-range-and-rms}}

Beker's original $rms_{2Hz}$ compare sites with the help of a single parameter\footnote{Beker originally defined $sigma\_{ET}$ also to distinguish the distributions of PSDs. In this paper we do not want to explore this quantity but focus only to the $rms$.}, using a particular frequency range from $2Hz$ to the upper frequency determined by the speed of the data acquisition. However, the noise budget of the low frequency part of ET is more frequency dependent. 

The term "seismic noise" covers two different aspects: the "original" seismic noise -- the movement of Earth shakes the mirrors -- and the Newtonian noise, or gravity gradient -- the seismic activity causes perturbation in the local gravity field. The first one can be damped by passive filtering (e.g. by a suspension system) but the second one cannot, thus active filtering is necessary \cite{Har15a,FioEta18a}. The seismic noise is relevant until  $1-2Hz$ and the Newtonian, or gravity gradient noise is relevant above that frequency up ot $7\,Hz$ according to ET low frequency sensitivity budget \cite{ETdes11r}. The exact values depend on the suspension system and the efficiency of the applied filtering methods.

\begin{figure}
	\centering
	\includegraphics[width=0.95\textwidth]{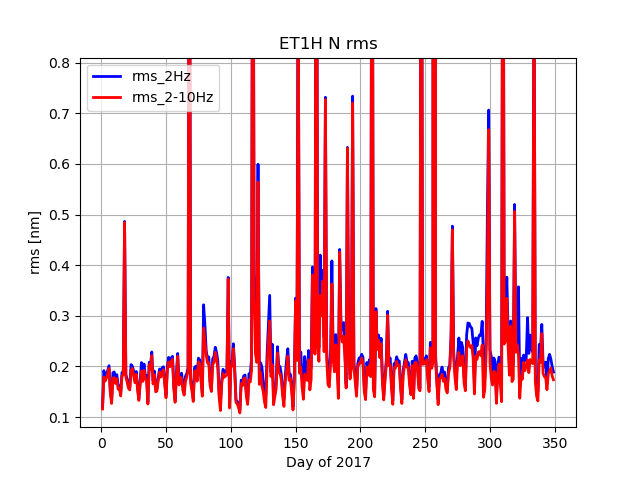}
	\includegraphics[width=0.95\textwidth]{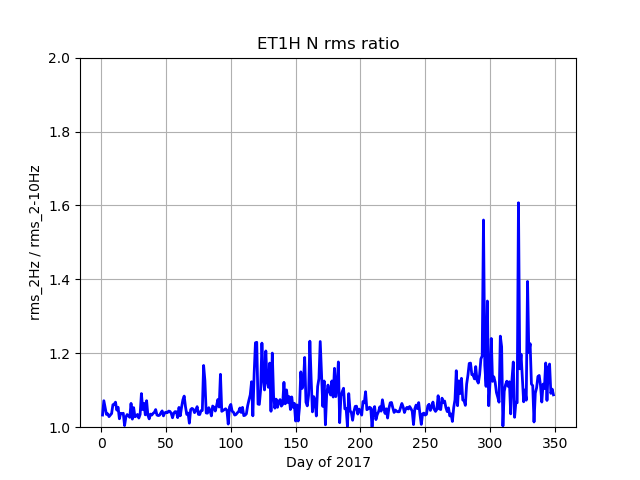}
	\caption{\label{fig:rmsOsszevetes}The upper figure displays the daily $rms_{2Hz}$ (blue) toghether with $rms_{2-10Hz}$ (red). The lower figure shows the ratio of the $rms_{2Hz}$ and $rms_{2-10Hz}$. The data is calculated from daily averages at the North direction of each day in 2017.}
\end{figure}

Therefore it is reasonable to consider the modification of the frequency range for cumulative characterisation of ET sites. There are two aspects that influence our choice. First, the rare very noisy events at high frequencies, above $10Hz$, are collected in $rms_{2Hz}$. This can be seen in Figure \ref{fig:rmsOsszevetes}.  Noises from this region eventuate irrelevant properties of the site, therefore high frequency cutoff is reasonable. 

\begin{figure}
	\centering
	\includegraphics[width=0.95\textwidth]{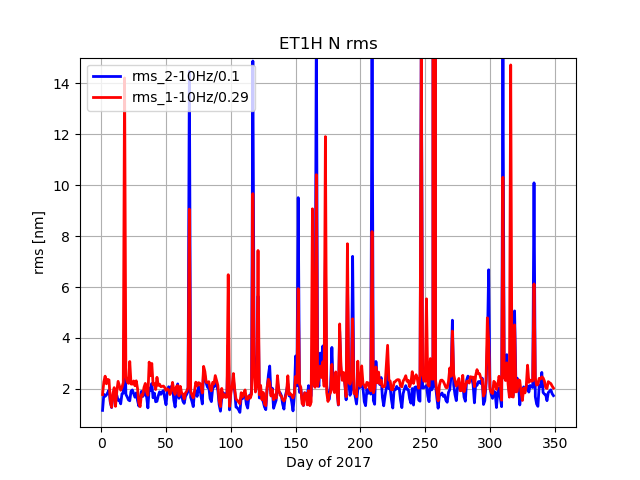}
	\includegraphics[width=0.95\textwidth]{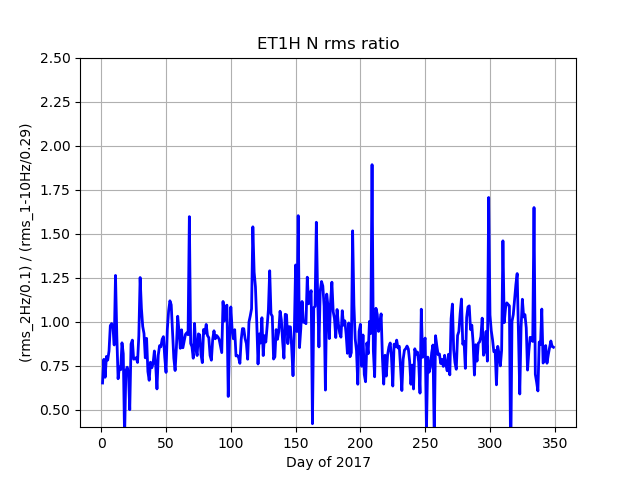}
	\caption{\label{fig:rmsOsszevetes1-10} The upper figure displays the normalized $rms_{2-10Hz}$ (blue) and $rms_{1-10Hz}$ values (red) of the daily averages in 2017 in the North direction. The lover figure shows the ratio of the same normalized $rms_{2-10Hz}$ and $rms_{1-10Hz}$. In both cases the normalization is made by the Black Forest values.}
\end{figure}

Also the low frequency limit is worth consideration. The recent cutoff at $2Hz$ was determined by the properties of the planned mirror suspensions. If one expects that mirror technology enables and science requires observations down to 1Hz, then the difference in the spectral properties of various sites must be characterized accordingly. Therefore we suggest to introduce suitable quantities and use also  $rms_{2-10Hz}$ and $rms_{1-10Hz}$ for further site selection information. The referential Black Forest line values are given in Eqs. \re{eq:rms1-10hz}. To illustrate it in Figure \ref{fig:rmsOsszevetes1-10} the normalized values of $rms$-s and they ratio are plotted. The figure indicates that there is a qualitative difference in the noisiness when lower frequencies are considered, otherwise one would expect an approximately constant ratio.

\section{Conclusions \label{sec:Conclusion2}}

In the previous sections we have studied and characterised the specific aspects of long term low frequency seismological data evaluation in order to find the best site characterisation measures for Einstein Telescope. Our general observation is, that there are several sensitive aspects in the spectral and cumulative characteristics and in their calculation methods. The differences may become significant when the noise spectra are  different and also, these performance measures are not the same from the point of view of potential ET requirements. 

In order to reduce this sensitivity we suggest the following improvements in site characterisation measures
\begin{enumerate}
	\item Use median instead of mode. Then we can omit the discretization and therefore its uncertainties and avoid STA sensitivity.  Also the mode is unstable if the distribution of the data contains new peaks, a phenomenon observed several times in our data. The use of median provides a selection.
	\item Use optimal STAs and INAs. That is advantageous for handling large amount of data. Moreover, the chosen interval length can be related directly to operational conditions and requirements of the low frequency part of the ET\footnote{The planned detection length of GW signals could reach 1-10ks. It could be reasonable to use a more suitable -- less than the order of expected detection length -- than the 14x128s averaging of Beker {\it et al} \cite{BekEta15a}. Furthermore the STA makes the overlapping much easier to handle.}. 
	\item Use both $rms_{2-10Hz}$ and $rms_{1-10Hz}$. The upper limit in the frequency range removes the information from the $rms$ which is irrelevant for the low frequency operation. The $1-10Hz$ frequency range enhances the lower frequency properties of the site.
\end{enumerate}

The suggested new $rms$ measures are different because of the mode-median difference and the change of the frequency range. We illustrate the differences in Table \ref{tab:rms_comp}, where the first row shows the reference values from the Black Forest line and the second row contains the values calculated from the 2017 data (349 days) of the ET1H station in the MGGL. Here the first column is calculated from the mode and the other columns from the median of the data.  The mode related and median related values calculated from the same data are different. Usually the median is larger, but not necessarily. A detailed evaluation of the MGGL data is shown in \cite{VanEta18m2}.

\begin{table}
\centering
\begin{tabular}{|c|c|c|c|c|}
\hline 
 & $rms_{2Hz}$ (mode) & $rms_{2Hz}$ &$rms_{2-10Hz}$ &  $rms_{1-10Hz}$ \tabularnewline
\hline 
\hline 
Black Forest line [nm] & 0.1 &  0.1 &  0.1 & 0.29 \tabularnewline
\hline 
ET1H 2017 [nm] & 0.136 &  0.153 & 0.152&  0.502 \tabularnewline\hline 
\end{tabular}
\caption{\label{tab:rms_comp} The various suggested $rms$ values for the Black Forest line and calculated from the 2017 data of the ET1H station.}
\end{table}

It is also remarkable that the expected duration of gravitational wave signals can be considered already in site selection. For example it may be reasonable to choose the STA periods according to observational requirements. If one expects, that a continuous observation of a minimal length (e.g. 128s for a black hole merger) is suitable, then the percentiles of low level averaging directly characterize the observation capabilities of the particular site.

We have seen that several seemingly minor aspects of the noise measures (e.g. the width of the noise levels in the mode calculations) may introduce different numbers and spectra, emphasizing different properties of the overall noisiness. Long periods are more sensitive to these aspects than short ones. 

\section{Acknowledgement}

The work was supported by the grants National Research, Development and Innovation Office \textendash{} NKFIH 116197(116375) NKFIH 124366(124508) and NKFIH 123815. The support of the PHAROS (CA16214) and G2net (CA17137) COST Actions is also acknowledged. The authors thank Géza Huba for the constant support and help, for Zoltán Zimborás and Jan Harms for important remarks. Also the help and support of the Nitrokémia Zrt and GEO-FABER Zrt. is greatly acknowledged.

\bibliographystyle{unsrt}

\begin{thebibliography}{1}

\bibitem{BraEta10p}
JFJ van~den Brand, MG~Beker, M~Doets, E~Hennes, and DS~Rabeling.
\newblock Einstein telescope site selection: {S}eismic and gravity gradient
  noise.
\newblock In {\em Journal of Physics: Conference Series}, volume 203, page
  012076. IOP Publishing, 2010.

\bibitem{Har15a}
Jan Harms.
\newblock Terrestrial gravity fluctuations.
\newblock {\em Living reviews in relativity}, 18(1):3, 2015.

\bibitem{ETdes11r}
ET~Science Team.
\newblock Einstein gravitational wave {T}elescope, {C}onceptual {D}esign
  {S}tudy.
\newblock Technical Report ET-0106C-10, June 2011.
\newblock http://www.et-gw.eu/etdsdocument.

\bibitem{BekEta15a}
M~G Beker, J~F~J van~den Brand, and D~S Rabeling.
\newblock Subterranean ground motion studies for the {E}instein {T}elescope.
\newblock {\em Classical and Quantum Gravity}, 32(2):025002, 2015.

\bibitem{Guralpmanual}
See the manual of the instrument: http://www.guralp.com/documents/DAS-030-0120.pdf


\bibitem{Bek13t}
M.~G. Beker.
\newblock {\em Low-frequency sensitivity of next generation gravitational wave
  detectors}.
\newblock PhD thesis, Vrije Universiteit Amsterdam, Amsterdam, June 2013.


\bibitem{BarEta17a}
G.G. Barnaf\"oldi, T.~Bulik, M.~Cieslar, E.~D\'avid, M.~Dobr\'oka, E.~Fenyvesi,
  Z.~Gr\'aczer, G.~Hamar, G.~Huba, \'A. Kis, R.~Kov\'acs, I.~Lemperger,
  P.~L\'evai, J.~Moln\'ar, D.~Nagy, A.~Nov\'ak, L.~Ol\'ah, P.~P\'azm\'andi,
  D.~Piri, D.~Rosinska, L.~Somlai, T.~Starecki, M.~Suchenek, G.~Sur\'anyi,
  S.~Szalai, D.~Varga, M.~Vas\'uth, P.~V\'an, B.~V\'as\'arhelyi, V.~Wesztergom,
  and Z.~W\'eber.
\newblock First report of long term measurements of the {MGGL} laboratory in
  the {M}\'atra mountain range.
\newblock {\em Classical and Quantum Gravity}, 34:114001(22), 2017.
\newblock arXiv: 1610.07630.



\bibitem{Pet93a}
J. Peterson.
\newblock {\em Observations and modeling of seismic background noise}.
Open-File Report,  USGS, 1993, RN:93-322.

\bibitem{FioEta18a}
D. Fiorucci, J. Harms, M. Barsuglia, I. Fiori, and F. Paoletti.
\newblock {Impact of infrasound atmospheric noise on gravity detectors used for astrophysical and geophysical applications}, 
{\em Physical Review D},  97(6):062003, 2018.

\bibitem{VanEta18m2}
P. Ván, et al.
\newblock {Long term measurements from Matra Gravitational and Geophysical Laboratory}, 
under publication

\end{thebibliography}

\end{document}